\documentstyle[12pt]{article}

\font\twlgot =eufm10 scaled \magstep1
\font\egtgot =eufm8
\font\sevgot =eufm7

\font\twlmsb =msbm10 scaled \magstep1
\font\egtmsb =msbm8
\font\sevmsb =msbm7

\newfam\gotfam
\def\pgot{\fam\gotfam\twlgot}
\textfont\gotfam\twlgot
\scriptfont\gotfam\egtgot
\scriptscriptfont\gotfam\sevgot
\def\got{\protect\pgot}

\newfam\msbfam
\textfont\msbfam\twlmsb
\scriptfont\msbfam\egtmsb
\scriptscriptfont\msbfam\sevmsb
\def\Bbb{\protect\pBbb}
\def\pBbb{\relax\ifmmode\expandafter\Bb\else\typeout{You cann't use
Bbb in text mode}\fi}
\def\Bb #1{{\fam\msbfam\relax#1}}

\def\thebibliography#1{\bigskip\section*{\large
\bf References\\}\list
  {[\arabic{enumi}]}{\settowidth\labelwidth{#1}\leftmargin\labelwidth
    \advance\leftmargin\labelsep
    \usecounter{enumi}}
    \def\newblock{\hskip .11em plus .33em minus .07em}
    \sloppy\clubpenalty4000\widowpenalty4000
    \sfcode`\.=1000\relax}

\def\op#1{\mathop{{\it\fam0} #1}\limits}

\newcommand{\id}{{\rm Id\,}}

\newcommand{\lng}{\langle}
\newcommand{\rng}{\rangle}

\newcommand{\beq}{\begin{equation}}
\newcommand{\eeq}{\end{equation}}
\newcommand{\ben}{\begin{eqnarray}}
\newcommand{\een}{\end{eqnarray}}
\newcommand{\be}{\begin{eqnarray*}}
\newcommand{\ee}{\end{eqnarray*}}
\newcommand{\bea}{\begin{eqalph}}
\newcommand{\eea}{\end{eqalph}}

\newcommand{\gG}{{\got G}}

\newcommand{\gE}{{\got E}}

\newcommand{\cK}{{\cal K}}

\newcommand{\ccG}{{\cal G}}

\newcommand{\la}{\lambda}

\newcommand{\f}{\phi}

\newcommand{\m}{\mu}

\newcommand{\g}{\gamma}

\newcommand{\ve}{\varepsilon}
\newcommand{\th}{\theta}

\newcommand{\si}{\sigma}

\newcommand{\bb}{{\bf 1}}

\newcommand{\wh}{\widehat}
\newcommand{\ol}{\overline}

\newcommand{\ot}{\otimes}

\newenvironment{eqalph}{\stepcounter{equation}
\setcounter{equationa}{\value{equation}}
\setcounter{equation}{0}

\begin{eqnarray}}{\end{eqnarray}\setcounter{equation}{\value{equationa}}}

\newcounter{example}
\newcounter{remark}
\newcounter{theorem}
\newcounter{proposition}
\newcounter{lemma}
\newcounter{corollary}
\newcounter{definition}

\def\theremark{\arabic{remark}}

\def\thedefinition{\arabic{definition}}

\newenvironment{prop}{\refstepcounter{definition} \medskip\noindent{\bf
Proposition \thedefinition.}}{\medskip }

\newcommand{\mar}[1]{}

\begin{document}
\hbox{}

{\parindent=0pt

{\large \bf Algebras of infinite qubit systems}  
\bigskip 

{\bf G. Sardanashvily}

\medskip

\begin{small}

Department of Theoretical Physics, Moscow State University, 117234
Moscow, Russia

E-mail: sard@grav.phys.msu.su

URL: http://webcenter.ru/$\sim$sardan/
\bigskip

{\bf Abstract.} The input and output algebras of an infinite qubit
system and their representations are described.

\end{small}
}

\bigskip
\bigskip

Let $Q$ be the two-dimensional complex space $\Bbb C^2$ equipped with
the standard positive non-degenerate Hermitian form $\lng.|.\rng_2$. 
Let $M_2$ be the algebra of complex $2\times 2$-matrices seen as a
$C^*$-algebra. A system of $m$ qubits is usually described by the
Hilbert space $E_m=\op\ot^m Q$ and the $C^*$-algebra $A_m=\op\ot^m M_2$, which
coincides with the algebra $B(E_m)$ of bounded operators in $E_m$. We 
straightforwardly generalize this description to an infinite
set $S$ of qubits by 
analogy with a spin lattice \cite{emch}. Its algebra $A_S$ admits
non-equivalent irreducible representations. If $S=\Bbb Z^+$, there is
one-to-one correspondence between the
representations of $A_S$ and those of an algebra of canonical
commutation relations. 

Given a system of $m$ qubits, one also considers an algebra of complex
functions on the set $\op\times^m\Bbb Z_2$. It is regarded as the
output algebra of a qubit system, while $A_m$ is the input one \cite{keyl}. 
There is a natural
monomorphism of this algebra to $A_m$. Using
the technique of groupoids \cite{rena}, we enlarge this algebra and
generalize it to infinite qubit systems.

We start from the input algebra, and follow the construction of
infinite tensor products of Hilbert spaces and 
$C^*$-algebras in \cite{emch}.
Let $\{Q_s,s\in S\}$ be a set of two-dimensional Hilbert spaces
$Q_s=\Bbb C^2$.
Let $\op\times_SE_s$ be the complex vector space whose elements are
finite linear combinations of elements $\{q_s\}$ of the Cartesian product
$\op\prod_S Q_s$ of the sets $Q_s$. The tensor product
$\op\ot_S Q_s$ of complex vector spaces $Q_s$ is the
quotient of $\op\times_S Q_s$ with respect to the vector subspace
generated by the elements of the form:
\begin{itemize}
\item $\{q_s\} + \{q'_s\} -\{q''_s\}$, where $q_r  + q'_r =q''_r$
for some element $r\in S$ and $q_s = q'_s =q''_s$ for all the others,
\item $\{q_s\} - \la\{q'_s\}$, $\la\in\Bbb C$, where $q_r= \la q'_r$
for some element $r\in S$ and $q_s = q'_s$ for all the others.
\end{itemize}
Given an element $\th=\{\th_s\}\in \op\prod_S Q_s$ such that all
$\th_s\neq 0$, let us denote 
$\ot^\th Q_s$ the subspace of 
$\op\ot_S Q_s$ spanned by the vectors $\ot q_s$ where $q_s\neq
\th_s$ only for a finite number of elements $s\in S$. It is called the
$\th$-tensor product of vector spaces $Q_s$, $s\in S$. 
Let us choose a family $\th=\{\th_s\}$ of normalized elements $\th_s\in
Q_s$, i.e., all $|\th_s|=1$.  Then $\ot^\th Q_s$ is a
pre-Hilbert space 
with respect to the positive non-degenerate Hermitian form
\be
\lng \ot^\th q_s|\ot^\th q'_s\rng:=\op\prod_{s\in S} \lng q_s|q'_s\rng_2.
\ee
Its completion $Q_S^\th$ is a Hilbert space whose orthonormal basis
consists of the elements $e_{ir}=\ot q_s$, $r\in S$, $i=1,2$, such that
$q_{s\neq r}=\th_s$ and $q_r=e_i$, where $\{e_i\}$ is an orthonormal
basis for $Q$.

Let now $\{A_s,s\in S\}$ be a set of unital $C^*$-algebras $A_s=M_2$. 
These algebras are provided with the operator norms
\be
\|a\|=(\la_0\ol\la_0 +\la_1\ol\la_1 +\la_2\ol\la_2 +\la_3\ol\la_3)^{1/2},
\qquad a=i\la_0\bb +\op\sum_{i=1,2,3}\la_i\si^i,
\ee
where $\si^i$ are the Pauli matrices.
Given the
family $\{\bb_s\}$, let us construct the $\{\bb_s\}$-tensor product
$\ot A_s$ of vector spaces $A_s$. It is a normed involutive algebra with
respect to the operations 
\be
(\ot a_s)(\ot a'_s)=\ot (a_sa'_s), \qquad (\ot a_s)^*=\ot a^*_s
\ee
and the norm 
\be
\|\ot a_s\|=\op\prod_s\|a_s\|.
\ee
Its completion $A_S$ is a $C^*$-algebra. Then the following holds \cite{emch}.

\begin{prop}
Given a family $\th=\{\th_s\}$ of normalized elements $\th_s\in Q_s$,
the natural representation of the involutive algebra 
$\ot A_s$ in the pre-Hilbert space $\ot^\th Q_s$ is extended to the
representation of the $C^*$-algebra $A_S$ in the Hilbert space
$Q_S^\th$ such that $A_S=B(Q_S^\th)$ is the algebra of all bounded
operators in $Q_S^\th$.
\end{prop}

\begin{prop}
Given two families $\th=\{\th_s\}$ and $\th'=\{\th'_s\}$ of normalized
elements, the representations of the $C^*$-algebra $A_S$ in the Hilbert
spaces $Q_S^\th$ and $Q_S^{\th'}$ are equivalent iff
\be
\op\sum_{s\in S} ||\lng\th_s|\th'_s|-1|<\infty.
\ee
\end{prop}

For instance, if $S=\Bbb Z^k$, we come to a spin lattice.

If $S$ is a countable set, let us consider its bijection onto $\Bbb
Z^+=\{1,2,\ldots\}$. Let us denote by $\si_r^i$, $r\in\Bbb Z^+$, the element
$\ot a_s\in A_S$ such that $a_r=\si^i$, $a_{s\neq r}=\bb$. Then the elements
\be
a_s^\mp=\frac12\si^3_1\cdots\si^3_{s-1}(\si^1_s \pm i\si^2_s)
\ee
make up the algebra $\ccG_S$ of canonical anticommutation
relations. The following holds \cite{emch}.

\begin{prop} There is one-to-one correspondence between the
representations $\pi$ of the $C^*$-algebra $A_S$ of a countable qubit
system and those $\pi^o$ of the CAR algebra $\ccG_S$, where $\pi^o$ is
the restriction of $\pi$ to $\ccG(S)\subset A_S$. Furthermore, a
representation $\pi$ is cyclic (resp. irreducible) iff $\pi^o$ is
cyclic (resp. irreducible). Representations $\pi$ and $\pi'$ of $A_S$
are equivalent iff its restrictions $\pi^o$ and $\pi'^o$ to $\ccG_S$ are
equivalent representations of $\ccG_S$. 
\end{prop}

Turn now to the output algebra. 
One can associate to a system of qubits $\{Q_s,s\in S\}$ the following
groupoid. Let $Z_2=\{\bb,\,p\,:\, p^2=\bb\}$ be the smallest Coxeter group.
Let us consider the set $X=Z_2{}^S$ of $Z_2$-valued functions on $S$. 
It is a set $2^S$ of all subsets of $S$, and it can be brought into a
Boolean algebra. 
Let $G\subset X$ be a subset of functions which equal $p\in Z_2$ 
at most finitely many points of $S$. Both $X$ and $G$ are commutative
Coxeter groups with respect to the pointwise multiplication. 
Given the action of $G$ on $X$ on the right, the product $\gG=X\times G$ 
can be brought into the action groupoid \cite{rena}, where: 
\begin{itemize}
\item a pair $((x,g),(x',g'))$ is composable iff $x'=xg$, 
\item the inversion $(x,g)^{-1}:=(xg,g^{-1})$,
\item the partial multiplication $(x,g)(xg,g')=(x,gg')$,
\item the range $r: (x,g)\mapsto (x,g)(x,g)^{-1}=(x,\bb)$,
\item the domain $l: (x,g)\mapsto (x,g)^{-1}(x,g)=(xg,\bb)$.
\end{itemize}
The unit space $\gG^0=r(\gG)=l(\gG)$ of this groupoid is naturally
identified with $X$. Given elements $x,y\in\gG^0$, 
let us denote the $r$-fibre
$\gG^x=r^{-1}(x)$, the $l$-fibre $\gG_y=l^{-1}(y)$, and
$\gG^x_y=\gG^x\cap\gG_y$.

Since $G$ acts freely on $X$, the action groupoid $\gG$ is principal,
i.e., the map 
$(r,l):\gG \to \gG^0\times \gG^0$ is an injection. 
Provided with the discrete topology, $X$ is
a locally compact space. Then $\gG$ is a locally compact groupoid.
Since $\gG^0\subset \gG$ is obviously an open subset, $\gG$ is an
$r$-discrete groupoid. 

Note that the action groupoid $\gG$ is isomorphic to the following one. 
Let us say that functions $x,y\in X$ are equivalent ($x\sim y$) iff
they differ  at most finitely many points of $S$. 
Let $\gG'\subset X\times X$ be the graphic of
this equivalence relation, i.e., it consists of pairs $(x,y)$ of
equivalent functions. 
One can provide $\gG'$ with the following groupoid
structure \cite{rena}: 
\begin{itemize}
\item a pair $((x,y),(y',z))$ is composable iff $y=y'$,
\item $(x,y)^{-1}:=(y,x)$,
\item $(x,y)(y,z)=(x,z)$,
\item $r: (x,y)\mapsto (x,x)$, $l:(x,y)\mapsto (y,y)$.
\end{itemize}
The unit space $\gG'^0$ of this groupoid is naturally
identified with $X$. The isomorphism of groupoids $\gG$ and $\gG'$ is
given by the assignment $\gG\ni (x,g)\mapsto (x,xg)\in \gG'$. In
particular, $\gG^x_y\neq\emptyset$ iff $x\sim y$.

Let $\cK(\gG,\Bbb C)$ be the space of complex functions on $\gG$ of
compact support provided with the inductive limit topology. Since $\gG$
is a discrete space, any function on $\gG$ is continuous. A left Haar
system for the groupoid $\gG$ is a family of measures 
$\{\m_x,\, x\in X\}$ on
$\gG$ indexed by points of the unit space $\gG^0=X$ such that:
\begin{itemize} 
\item the support of the measure $\m_x$ is $\gG^x$,
\item for any $(x,g)\in\gG$ and any $f\in\cK(\gG,\Bbb C)$, we have
\be
\int f((x,g)(y,g')) d\m_{xg}((y,g'))=\int f((y,g'))d\m_x((y,g')).
\ee
\end{itemize}
A left Haar system
for the action groupoid $\gG$ exists. It is given by the measures
$\m_x=\ve_x\times \m_G$, where $\ve_x$ is the Dirac measure on $X$ with
support at $x\in X$ and $\m_G$ is the left Haar measure on the locally
compact group $G$. We have
\be
\int f((y,g))d\m_x((y,g))=\int f((x,g))d\m_G(g)=\op\sum_{g\in G}f((x,g)),
\ee
where the sum is finite since $f\in\cK(\gG,\Bbb C)$ are of compact support.
 
With a left Haar system, the space $\cK(\gG, \Bbb C)$ is brought into
an involutive algebra as follows \cite{rena}. Since the
$U(1)$-extension of $G$ is 
trivial up to equivalence, the algebraic operations in $\cK(\gG,
\Bbb C)$ read 
\mar{q1,2}\ben
&& (f*f')((x,g))=\op\sum_{g'\in G} f((x,gg'))f'(xgg'^{-1},g'^{-1}),
\label{q1} \\
&& f^*((x,g))=\ol{f((x,g)^{-1})}. \label{q2}
\een
This algebra can be provided with the norm
\mar{q3}\beq
\|f\|=\max(\op\sup_x \op\sum_g|f(x,g)|,\,
\op\sup_x \op\sum_g|f(xg,g^{-1})|). \label{q3}
\eeq

In particular, let us consider a subspace $A_G\subset \cK(\gG, \Bbb C)$
of functions $f((x,g))=f(g)$ independent of $x\in X$. The algebraic
operations (\ref{q1}) -- (\ref{q2}) on these functions read
\be
(f*f')(g)=\op\sum_{g'\in G} f(gg')f'(g'^{-1}), \qquad
f^*(g)=\ol{f(g^{-1})}=\ol{f(g)}.
\ee
Thus, $A_G$ is exactly the group algebra of the locally compact group
$G$ provided with the norm
\be
\|f\|= \op\sum_g|f(g)|.
\ee 
There is the monomorphism of this algebra to the algebra $\ot A_s$ as follows.
Let us assign to each element $g\in G$ the element $\wh g=\ot a_s\in \ot
A_s$, where $a_s=\bb$ if $g(s)=\bb$ and $a_s=\si^1$ if $g(s)=p$. Then
the above mentioned monomorphism is given by the association 
\be
f(g)\mapsto \op\sum_g f(g)\wh g.
\ee
Thus, one can think of $A_G$ and, consequently, 
$\cK(\gG, \Bbb C)$ as being a generalization of the output
algebra in \cite{keyl}. 

The monomorphism $A_G\to \ot A_s$ provides the representations of the 
involutive algebra $A_G$ in Hilbert spaces $Q_S^\th$.
One can construct representations of the whole algebra $\cK(\gG, \Bbb
C)$ as follows. 

Given the Hilbert space $Q_S^\th$, let us consider the
product $\gE=X\times Q_S^\th$ seen as a group bundle in the Abelian groups 
$Q_S^\th$ over $X$. It is a groupoid such that
\begin{itemize}
\item a pair $((x,v), (x',v'))$, $x,x'\in X$, $v,v'\in Q_S^\th$, is
composable iff $x'=x$,  
\item $(x,v)^{-1}:=(x,-v)$,
\item $(x,v)(x,v')=(x,v+v')$,
\item $r((x,v))=l((x,v))=x$.
\end{itemize}
Its unit space is $X$, and its fibres $\gE_x$ are isomorphic to
$Q_S^\th$. Let Iso$\,\gE$ denote the set of all isomorphisms 
$\g^x_y:\gE_y\to \gE_x$, $x,y\in \gG^0$. This set possesses the natural
groupoid structure such that:
\begin{itemize}
\item a pair $(\g^x_y,\g^{x'}_{y'})$ is composable iff $x'=y$,
\item $(\g^x_y)^{-1}$ is the inverse of $\g^x_y$,
\item $\g^x_y\g^y_z=\g^x_y\circ \g^y_z$,
\item $r(\g^x_y)=\id\gE_x$ and $l(\g^x_y)=\id\gE_y$.
\end{itemize}
The unit space $\gE^0$ of this groupoid is naturally identified with
$X$ by the association $x\mapsto \id\gE_x$. 
The assignment  
\be
(x,g) \mapsto \g^{xg}_x=\wh g
\ee
provides a homomorphism of the action groupoid $\gG$ to Iso$\,\gE$ and,
consequently, a representation of $\gG$ in $\gE$.

Let the discrete topological space $X$ be provided with the measure 
$\m_X$ such that
\be
\int f(x)d\mu_X(x)=\op\sum_x f(x), \qquad f\in \cK(X,\Bbb C).
\ee
Let $L^2(\gE,\m_X)$ denote the Hilbert space of square $\m_X$-integrable
sections $\f$ of the group bundle $\gE\to X$. Then we have a desired
representation $\pi$ of the algebra $\cK(\gG, \Bbb C)$ in 
$L^2(\gE,\m_X)$ by the formula
\be
(\pi(f)\f)(x)=\op\sum_g f((xg^{-1},g))\wh g \f(xg^{-1}).
\ee
The algebra $\cK(\gG, \Bbb C)$ can be provided with the operator norm
$\|f\|=\|\pi(f)\|$ with respect to this representation. Its completion
relative to this norm is the $C^*$-algebra of the action groupoid $\gG$.


\begin{thebibliography}{ederf}

\bibitem{emch} G.Emch, {\it Algebraic Methods in Statistical Mechanics
and Quantum Field Theory} (Wiley--Interscience, New York, 1972).

\bibitem{keyl} M.Keyl, Fundamentals of quantum information theory, {\it
Phys. Rep.} {\bf 369} (2002) 431.

\bibitem{rena} J.Renault, {\it A Groupoid Approach to $C^*$-algebras},
Lect. Notes Math. {\bf 793} (Springer-Verlag, Berlin, 1980).


\end{thebibliography}
\end{document}